\documentclass{llncs}

\usepackage{color}
\usepackage[usenames,dvipsnames,svgnames,table]{xcolor}

\usepackage[titletoc,toc,title]{appendix}
\usepackage{float}
\usepackage{placeins}
\usepackage{graphicx}
\usepackage{caption}
\usepackage{pbox}
\usepackage{rotating}
\usepackage{pdflscape}

\usepackage{subcaption}
\usepackage{tikz}
\usetikzlibrary{arrows}

\usepackage{adjustbox}
\usepackage{amsmath}
\usepackage{chapterbib}

\makeatletter
\def\@seccntformat#1{\@ifundefined{#1@cntformat}%
   {\csname the#1\endcsname\quad}  
   {\csname #1@cntformat\endcsname}
}
\let\oldappendix\appendix 
\renewcommand\appendix{%
    \oldappendix
    \newcommand{\section@cntformat}{\appendixname~\thesection\quad}
}
\makeatother

\usepackage{hyperref}
\hypersetup{
  linkbordercolor=red!80!black,
	citebordercolor=green!80!black,%
  urlcolor={blue!80!black}
}

\begin{document}

\title{Twitter Sentiment Analysis Applied to Finance: A Case Study in the Retail Industry}
\titlerunning{Twitter Sentiment Analysis Applied to Finance - A Case Study in the Retail Industry}  
%
\author{Th\'{a}rsis T. P. Souza\inst{1}, Olga Kolchyna\inst{1}, Philip C. Treleaven\inst{1}\inst{2} \and Tomaso Aste\inst{1}\inst{2}}

\authorrunning{Souza et al.} 
%
\tocauthor{Souza et al.}
\institute{Department of Computer Science, UCL, Gower Street, London, UK,\and
Systemic Risk Centre, London School of Economics and Political Sciences, London, UK}

\maketitle              

\begin{abstract}
This paper presents a financial analysis over Twitter sentiment analytics extracted from listed retail brands. We investigate whether there is statistically-significant information between the Twitter sentiment and volume, and stock returns and volatility. Traditional newswires are also considered as a proxy for the market sentiment for comparative purpose. The results suggest that social media is indeed a valuable source in the analysis of the financial dynamics in the retail sector even when compared to mainstream news such as the Wall Street Journal and Dow Jones Newswires.
\\
\keywords{retail, financial markets, data science, computational social science, social media, news analytics}
\end{abstract}

%
%

\section{Introduction}

Major news announcements can have a high impact on the financial market and investor behaviour resulting in rapid changes or abnormal effects in financial portfolios. As human responsiveness is limited, automated news analysis has been developed as a fundamental component to algorithmic trading. In this way, traders can shorten the time of reaction in response to breaking stories. The basic idea behind these news analytics technologies is to predict human behavior and automate it, so traders may be able to anticipate asset movements before making an investment or risk management decision.

Twitter data has also become an increasingly important source to describe financial dynamics. It provides a fine-grained real-time information channel that includes not only major news stories but also minor events that, if properly modelled, can provide ex-ante information about the market even before the main newswires. Recent developments have reflected this prominent role of social media in the financial markets. One major example is the U.S. Securities and Exchange Commission report allowing companies to use Twitter to announce key information in compliance with Regulation Fair Disclosure \cite{SEC:2013}. Twitter has also shown that can cause fast and drastic impact. In 2013, with the so-called \textit{Hash Crash}, a hacked Twitter account of the American news agency Associated Press falsely disclosed a message about an attack on the White House causing a drop in the Dow Jones Industrial Average of 145 points in minutes \cite{WSJ:2013}.

In \cite{1507.00955} we proposed a new model for sentiment classification using Twitter. We combined the traditional lexicon approach with a support vector machine algorithm to achieve better predictive performance. In the present work, we use the dataset of sentiment analytics from \cite{1507.00955} to investigate the interplay between the Twitter sentiment extracted from listed retail brands, and financial stock returns and volatility. We verify whether there is statistically-significant information in this relationship and also compare it to a corresponding analysis using sentiment from traditional newswires. We consider volatility and log-returns as financial endogenous variables and we take Twitter sentiment and volume as exogenous explanatory variables in the financial dynamics of the selected stocks. We also consider traditional newswires as datasource for comparative purpose. Therefore, the main objectives are: (i) verify whether there is statistically-significant information between the Twitter sentiment, and stock returns and volatility and (ii) compare this interplay while using mainstream news as a proxy for the market sentiment. The main contribution of this work is an empirical evidence that supports the use of Twitter as a significant datasource in the context of financial markets in the retail industry even when compared to traditional newswires.

\section{Literature Review}

The investigation of the market impact of News has been long studied since the seminal work of \cite{cutler1989moves}. In this work, the authors investigate to which extent macroeconomic news explain return variance and also analyse the observed market moves following major political and world events. More recently, \cite{tetlock2007giving} provided the first evidence that news media content can predict movements in broad indicators of stock market activity. The authors found correlation between high/low pessimism of media and high market trading volume. They further analyse the relation between the sentiment of news, earnings and return predictability \cite{tetlock2008more}. Since then, with the availability of machine readable news and the use of sentiment analysis, several works have found news as a significant source for financial applications: \cite{Tobias:2013} found positive correlation between the number of mentions of a company in the Financial Times and its stock's volume; \cite{Lillo2012} investigate the effect of News in the behavior of the traders; \cite{Mitra:2013} analyse the Thomson Reuters News Analytics (TRNA) and find a causality between sentiment and, volatility and liquidity. 

Recent research supports the hypothesis that Twitter data also has statistically-significant information related to financial indicators. As one of the first investigations analysing Twitter in the context of financial markets, \cite{Bollen20111} analyse the text content of daily Twitter feeds to identify two types of mood: (i) polarity (positive vs. negative) and (ii) emotions (calm, alert, sure, vital, kind, and happy). They were able to increase the accuracy in the prediction of the DJIA index. Similar work \cite{Zhang_tradingstrategies} was able to predict not only the DJIA index but also the NASDAQ-100 index; the authors measured the agreement of sentiment between messages in addition to the market mood. More recently, \cite{citeulike:13108056} combined Information Theory with sentiment analysis to demonstrate that Twitter sentiment can contain statistically-significant ex-ante information on future prices of the S\&P500 index and also identified a subset of securities in which hourly changes in social media sentiment do provide lead-time information. As a contribution to the to the field of event study research, \cite{Timm:2014} offer a methodology to analyse market reactions to combinations of different types of news events using Twitter to identify which news are more important from the investor perspective. In a similar way, \cite{1506.02431} combine sentiment analytics with the identification of Twitter peaks in an event study approach to imply directions of market evolution. Further, exploring the social network structure from Twitter users, \cite{YangTwitter:2014} provide empirical evidence of a financial community in Twitter in which users' interests align with the financial market.

Similar to the present work, \cite{Smailovic:2013} investigate the causality between polarity measures from Twitter and daily return of closing prices. The authors also use sentiment derived from a Support Vector Machine model to classify the tweets into the positive, negative and neutral categories. As a contribution compared to this work we not only investigate the causality in returns but also stock's volatility. Also, we provide a comparison between Twitter and traditional newswires. Moreover, we concentrate the analysis in retails brands which can provide meaningful insights for applications in that domain.

Further examples of social media applications in the stock market are: the use of StockTwits sentiment and posting volume
to predict daily returns, volatility and trading volume \cite{Oliveira:2013}; the extraction of features from financial message board for stock market predictions \cite{Sehgal:2007:SSP:1335998.1336036} and approaches combining Twitter with other sources such as blogs and news \cite{Zhang_tradingstrategies, 2014arXiv1405.3117S, 6924062}.

\section{Dataset}

Our analysis is conducted on a set of five listed retail brands with stocks traded in the US equity market, which we
monitor during the period from November 01, 2013 to September 30, 2014. The name of the investigated stocks with respective Reuters Instrument Codes (RIC) follow: ABERCROMBIE \& FITCH CO. (ANF.N), NIKE INC. (NKE.N), HOME DEPOT INC. (HD.N), MATTEL INC. (MAT.N) and GAMESTOP CORP. (GME.N). The choice of companies is bounded by the Twitter sentiment analytics dataset provided by \cite{1507.00955}.

Given the companies selected, we consider three streams of time series data: (i) the market data, which is given at the daily stock price; (ii) news metadata supplied by \cite{RAVENPACK}, which consists in 10,949 news stories from Dow Jones Newswires, the Wall Street Journal and Barron's, and (iii) the social media data analytics provided by \cite{1507.00955}, which is based on 42,803,225 Twitter messages.

\subsection{News Analytics}
The news  analytics  data  supplied by \cite{RAVENPACK} are  provided  in  a  metadata  format  where each news receives scores quantifying  characteristics such as relevance and sentiment according to a related individual stock. Table \ref{tb:NA} shows a sample of the news sentiment analytics data provided. The relevance score ($Relevance$) of the news ranges between 0 and 100 and indicates how strongly related the company is to the underlying news story, with higher values indicating greater relevance. Usually, a relevance value of at least 90 indicates that the entity is referenced in the main title or headline of the news item, while lower values indicate references further down the story body. Here we filter the news stories with a relevance of 100. This increases the likelihood of the story considered being related to the underlying equity. Besides the relevance, we also consider the Event Sentiment Score ($ESS$). This measure indicates short-term positive or negative financial or economic impact of the news in the underlying company; higher values indicate more positive impact. It ranges between 0 and 100 where higher values indicate more positive sentiment while lower values below 50 show negative sentiment.

\begin{table}[H]
\centering
\caption{News Sentiment Analytics. Each line represents a news story related to a company. The metadata considered consists of the relevance and sentiment scores and a timestamp.}
\label{tb:NA}
\begin{center}
\begin{adjustbox}{max width=\textwidth}
 \begin{tabular}{l l c c c c} 
\hline
 Story & Company & Date & Hour & Relevance & Event Sentiment Score (ESS)\\
 \hline
1 & NIKE INC. & 20140104 &	210130 &	33 &	64 \\

2 & MATTEL INC. & 20140105 &	41357 &	100 &	50 \\
 
3 & NIKE INC. & 20140105 &	145917 &	93 &	88\\
 
4 & NIKE INC. & 20140105 &	150523 &	100 &	61\\
 
5 & GAMESTOP CORP. & 20140105 &	193507 &	44 &	50\\

6 & GAMESTOP CORP. & 20140106 &	170040 &	99 &	44\\
 
7 & MATTEL INC. & 20140106 &	222532 &	100 &	61\\
 
8 & GAMESTOP CORP. & 20140107 &	32601 &	100 &	50\\
 
9 & MATTEL INC.& 20140107 &	172628 &	55 &	40\\
 
10 & NIKE INC. & 20140110 &	204027 &	100 &	67\\
 \hline
\end{tabular}
\end{adjustbox}
\end{center}
 \end{table}  

Given this metadata information, we first normalize the Event Sentiment Score ($ESS$) of a given story at a timestamp $\Delta t$ such as it ranges between -1 and 1, and we label it as $\widehat{ESS}(\Delta t) \in \left[-1,1\right]$. Then, we define the sentiment and volume analytics for each company as:
\begin{itemize}
	\item $G_{News}(t)$: daily number of positive News, i.e., daily total number of News with $\widehat{ESS}(\Delta t)  > 0$; \\
	\item $B_{News}(t)$: daily number of negative News, i.e., daily total number of News with $\widehat{ESS}(\Delta t)  < 0$; \\
	\item $V_{News}(t)$: daily total number of News; \\
	\item $SA_{News}(t)$: daily absolute sentiment from News: 
	\begin{equation}
SA_{News}(t) = G_{News}(t) - B_{News}(t);
\end{equation}
	\item $SR_{News}(t) \in \left[-1,1\right]$: daily relative sentiment from News as the daily mean of sentiment score $\widehat{ESS}(\Delta t)$, $\Delta t \in \left[t,t+1\right)$.
\end{itemize}

\subsection{Twitter Analytics}

For the Twitter data analytics, we use the dataset from \cite{1507.00955}. It provides sentiment and volume metrics related to a company. We use the following analytics:
\begin{itemize}
 \item $G_{Twitter}(t)$: daily number of positive English tweets;\\
	\item $B_{Twitter}(t)$: daily number of negative English tweets;\\
	\item $V_{Twitter}(t)$: daily total number of messages regardless of the language.
\end{itemize}

Table \ref{tb:TA} shows an example of the Twitter sentiment analytics for the company MATTEL INC. For the  polarity classification, \cite{1507.00955} employed a new approach based on the combination of existing common used techniques (lexicon-based and machine learning based) which outperformed standard benchmarks, see \cite{1507.00955} for further details. Notice that the number of positive, negative and neutral messages do not sum up to the total volume, as the former consider only English tweets and the total volume covers the total number of messages regardless of the language. Also, although provided, we do not use the number of neutral messages as we believe that the extreme polarities (positive and negative) may be more informative.

  \begin{table}[H]
\centering
\caption{Twitter Sentiment Analytics. Sample of analytics for the company MATTEL INC. It shows the positive, negative and neutral English Twitter messages related to the company and also the total number of messages regardless of the language.}
\label{tb:TA}
\begin{center}
 \begin{adjustbox}{max width=\textwidth}
 \begin{tabular}{c c c c c c} 
\hline
 Date & CompanyID & Volume & \#Positive & \#Negative & \#Neutral \\
 \hline
01/11/2013& MATTEL INC. & 1,980 & 8 & 4 & 485 \\
 
02/11/2013& MATTEL INC. & 1,750 & 12 & 2 & 339\\
 
03/11/2013& MATTEL INC. & 1,700 & 8 & 1 & 518\\
 
04/11/2013& MATTEL INC. & 2,720 & 19 & 2 & 429\\

05/11/2013& MATTEL INC. & 1,980 & 11 & 8 & 793\\

06/11/2013& MATTEL INC. & 1,580 & 11 & 4 & 470\\
 
07/11/2013& MATTEL INC. & 1,770 & 7 & 1 & 498\\
08/11/2013& MATTEL INC. & 1,900 & 5 & 4 & 288\\
09/11/2013& MATTEL INC. & 1,260 & 16 & 2 & 236\\
10/11/2013& MATTEL INC. & 1,700 & 7 & 8 & 313\\
 \hline
\end{tabular}
\end{adjustbox}
\end{center}
 \end{table}  \FloatBarrier

We hence compute the variables:
\begin{itemize}
	\item $SA_{Twitter}(t)$: daily absolute sentiment from Twitter: 
		\begin{equation}
SA_{Twitter}(t) = G_{Twitter}(t) - B_{Twitter}(t);
\end{equation}
	\item $SR_{Twitter}(t) \in \left[-1,1\right]$: daily relative sentiment from Twitter as 
	\begin{equation}
	SR_{Twitter}(t) = \frac{G_{Twitter}(t) - B_{Twitter}(t)}{G_{Twitter}(t) + B_{Twitter}(t)}.
	\end{equation}
\end{itemize}
Notice that $SR_{Twitter}(t_0) = +1$, represents a day $t_0$ with the highest positive sentiment for the company considered; conversely $SR_{Twitter}(t_0) = -1$ indicates the highest negative sentiment, whereas we consider neutrality when $SR_{Twitter}(t_0) = 0$.

Although Twitter and News are distinct datasources, noticed that we have computed sentiment and volume analytics in such way that we have comparable time series between those sources\footnote{We may refer to a time series independently to a specific datasource, in such cases we will represent it as its original symbol but without the text subscript, e.g., the number of positive will be represented as $G(t)$ when discussing both News $G_{News}(t)$ and Twitter $G_{Twitter}(t)$ in the same context.}. This allows us make a comparative study between them while analysing the financial data further defined. 

Table \ref{tb:Comp} shows a summary description of the selected companies with the number of stories considered. We show the total number of News related to the each company and also the number of relevant news, i.e., those in which the news story has a $Relevance$ score equals to 100, as explained previously. Moreover, we present the total number of tweets related to each company. Notice that the Twitter dataset used does not provide any relevance score, therefore there is no further filtering process.

  \begin{table}[H]
	\centering
\caption{Summary table of selected companies. The five retails brands selected for the analysis along with their market capitalization. Also, we present the total number of news and tweets in the selected period. The relevant news represent the news filtered with the highest relevance score (100).}
\label{tb:Comp}
\begin{center}
 \begin{adjustbox}{max width=\textwidth}
 \begin{tabular}{l c c c c c} 
\hline
Company & RIC	& Market Cap.* (\$Billions)	& Total No. of News	& Relevant News & No. of Tweets	\\ \hline
ABERCROMBIE \& FITCH CO. & ANF.N &  2.86 & 1,608 & 174 & 1,352,643\\ 
NIKE INC. & NKE.N & 67.39 & 2,881 & 178 & 38,033,900\\
HOME DEPOT INC. & HD.N & 111.57 & 3,835 & 241 & 1,593,204\\ 
MATTEL INC. & MAT.N &  15.02  & 1,508 & 125  & 613,798\\
GAMESTOP CORP. & GME.N &  6.41  & 1,117 & 167 & 1,209,680
\\
\hline
\multicolumn{6}{l}{(*) Market Capitalization as in October 31, 2013. Source: Thomson Reuters Ikon.} \\
\end{tabular}
\end{adjustbox}
\end{center}
 \end{table}  \FloatBarrier

\section{Financial Variables}

Let $P(t)$ be the closing price of an asset at day $t$ and $R(t) = \log{P(t)} - \log{P(t-1)}$ its daily log-return. We consider the Excess of Log-return\footnote{An alternative approach is to examine the alpha generation as the excess return of the underlying stock relative to its benchmark adjusted for a given level of risk as in the market model described in \cite{JOFI:JOFI518}.} of the asset over the return of the market index $\widehat{R}$ as:
\begin{equation}
\label{eq:ERDef}
ER(t) = R(t) - \widehat{R}(t).
\end{equation}
We consider the S\&P500 daily returns as the market index $\widehat{R}$.

As a proxy for the daily volatility of a stock, we define:
\begin{align}
VOL(t) = 2\frac{P_{high}(t) - P_{low}(t)}{P_{high}(t) + P_{low}(t)},
\end{align}
where $P_{high}(t)$ and $P_{low}(t)$ are the highest and the lowest price of the stock at day $t$, respectively.

\begin{figure}[!h]
\centering
\scalebox{0.35}{\includegraphics{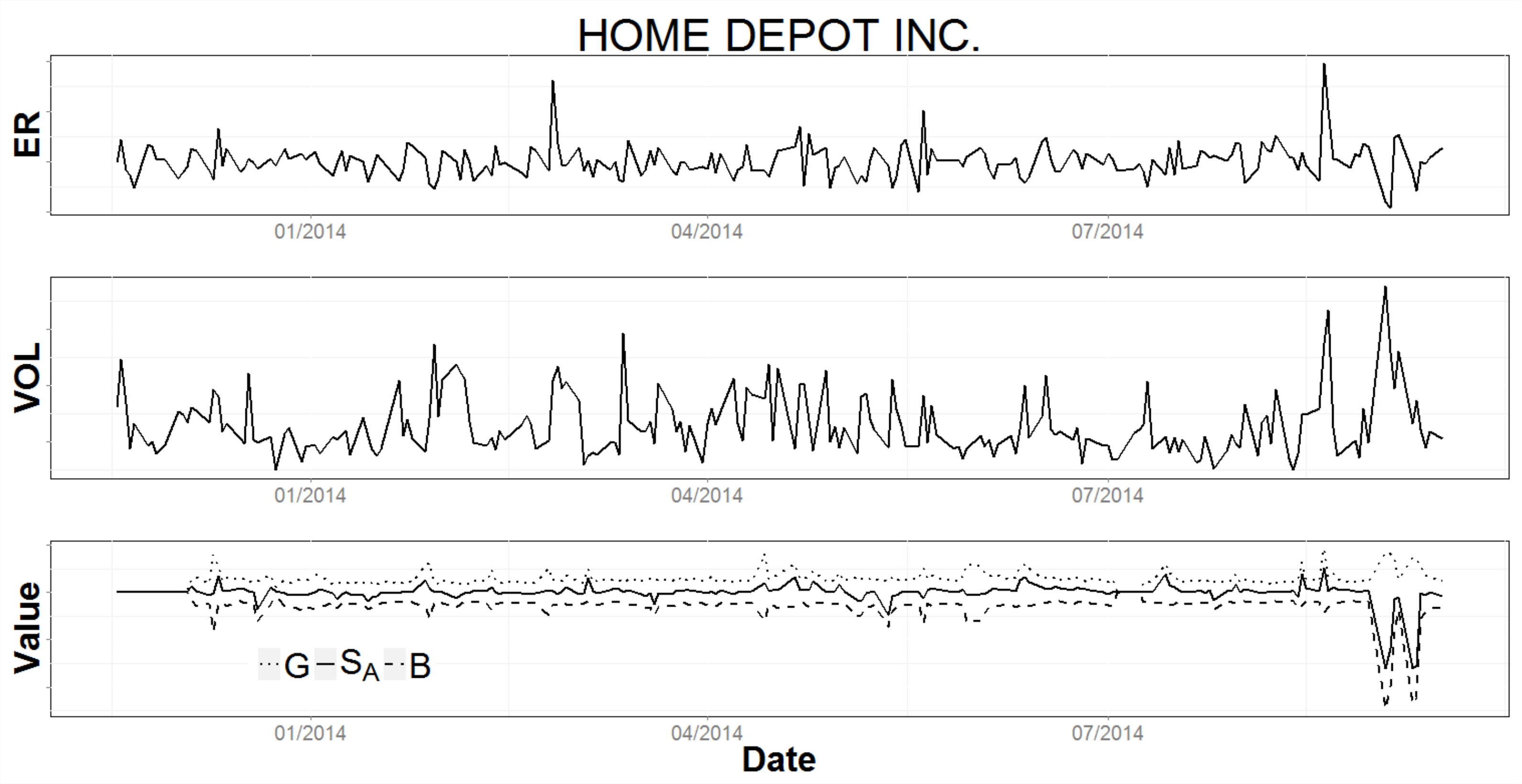}}
\caption{Twitter's Sample Descriptive Analysis for Home-Depot Inc. Variables: Excess of log-return, $ER$; volatility, $VOL$; absolute sentiment, $SA_{Twitter}$; number of positive messages, $G_{Twitter}$ and number of negative messages, $B_{Twitter}$. There is a spike in volume and decrease in sentiment at the end of the period which follows a corresponding drop in excess of log-return.}
	\label{fig:HD-Descriptive-Twitter}
 \end{figure}

\begin{figure}[t]
\centering
\scalebox{0.17}{\includegraphics{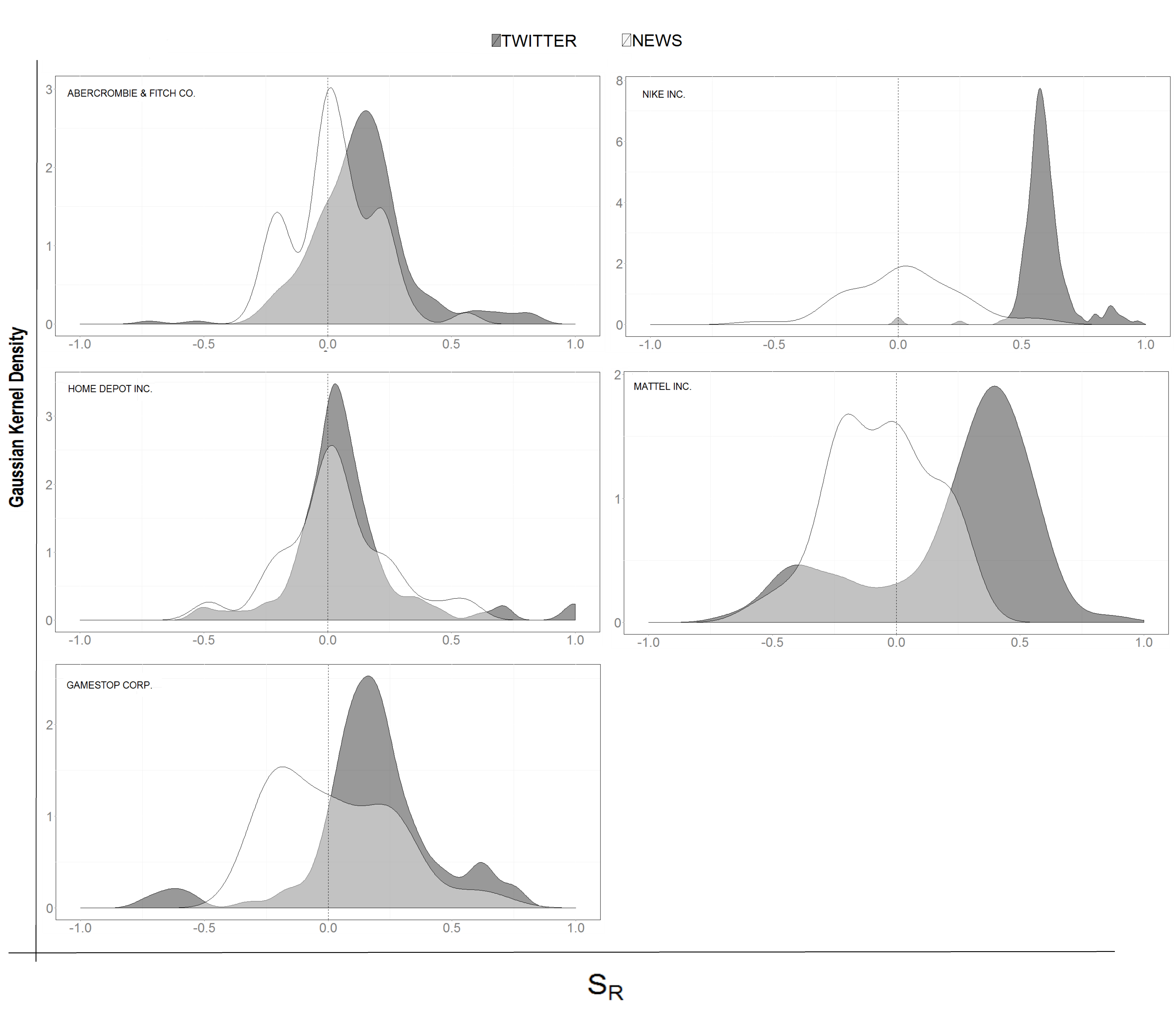}}
\caption{Distribution of relative sentiment from Twitter $SR_{Twitter}(t)$ and News $SR_{News}(t)$ for the companies: ABERCROMBIE \& FITCH CO., GAMESTOP CORP., HOME DEPOT INC., MATTEL INC. and NIKE INC. It is clear that the sentiment provided by Twitter is a distinct proxy for market sentiment compared to News as each company analysed depicts different distributions of sentiment.}
\label{fig:descriptive-SR}
 \end{figure}

Fig. \ref{fig:HD-Descriptive-Twitter} shows a sample of the calculated variables from Twitter for the company Home-Depot Inc. It is interesting to notice a spike in volume and decrease in sentiment at the end of the period which follows a corresponding drop in excess of log-return. Further, Fig. \ref{fig:descriptive-SR} depicts  the distribution of values of the relative sentiment obtained from Twitter and News. We observe that both present a skewed distribution while news has a more neutral-centred distribution compared to Twitter. It is important to notice that the sentiment provided by the Twitter analytics presents a distinct proxy for sentiment compared to News as each company analysed depicts different positive/negative sentiment tones, e.g., the NIKE's Twitter sentiment is highly positive while the news' sentiment has a mean around a neutral point.

    \FloatBarrier

    \FloatBarrier

\section{Method}

\subsection{Granger Causality}
We are interested in investigating the statistical causality between sentiment and the financial variables. In this sense, \cite{granger1980testing} introduced a concept of cause-effect dependence where the cause not only should occur before the effect but also should contain unique information about the effect. Therefore, we say that X Granger-cause Y if the prediction of Y can be improved using both information from X and Y as compared to only utilizing Y.

In a Vector Auto-Regressive (VAR) framework, we can assess the Granger causality performing a F-test to verify the null hypothesis that Y is not Granger-caused by X and measure its probability of rejection within a confidence level. Hence, assuming the VAR models:
\begin{align}
Y_{t} &= \alpha_0 + \alpha_1 Y_{t-1} + \ldots +  \alpha_k Y_{t-k} + \beta_1 X_{t-1} + \ldots +  \beta_k X_{t-k} +\epsilon_t  \label{eq:VAR1},\\
X_{t} &= \gamma_0 + \gamma_1 X_{t-1} + \ldots +  \gamma_k X_{t-k} + \theta_1 Y_{t-1} + \ldots +  \theta_k Y_{t-k} +\widehat{\epsilon}_t  \label{eq:VAR2},
\end{align}
we take the null hypothesis in equation \eqref{eq:H1} and test it against its alternative one in equation \eqref{eq:H2}. Thus, a rejection of the null hypothesis implies that Y Granger-cause X.
\begin{align}
\mathcal{H}_0&: \beta_1 = \beta_2 = \ldots = \beta_k = 0 \label{eq:H1}\\
\mathcal{H}_1&:\exists \beta_{\tau}, \ 0 \leq \tau \leq k  : \ \ \beta_{\tau} \neq 0 \label{eq:H2}
\end{align}
In the same way, the test for X Granger-cause Y can be done considering the equation \eqref{eq:VAR2} and taking the hypotheses from equations \eqref{eq:H1} and \eqref{eq:H2} in an analogous way.

For both News and Twitter, we will test the Grange-causality between the Excess of Log-return $ER$ and the number of positive stories $G$, the number of negative stories $B$, the relative sentiment $SR$ and the absolute sentiment $SA$. For the volatility $VOL$, we will consider the total volume of stories $V$ in addition to previously mentioned variables. Furthermore, we will perform the Granger-causality test over the normally standardized versions of the time series analysed such as they have zero mean and standard deviation 1. To perform the F-statistics of the Granger-causality tests we use the function anova.lm from the package stats of the R Project for Statistical Computing \cite{R2015}.

For the only purpose of visualization of the Granger-causality results we create a Granger-causality graph $G = \left[V, E\right]$, where $V$ is a node set, and $E$ is an edge set. A node $u \in V$ represents a variable in the causality test and an edge $e = \left(u, v\right)$ indicates that $u$ Granger-causes $v$ within a pre-defined significance level. Further, we define $p(e)$ as an attribute of the edge and if $C$ is the set of companies in which we have a causality between $u$ and $v$, then we set $p(e) = C$. Fig. \ref{fig:G} shows an example of a Granger-causality graph that indicates that $u$ Granger-causes $v$ for the set of companies $C$.

\tikzstyle{central_node_fill}=[circle,fill=black!20,draw,font=\sffamily\Large\bfseries]

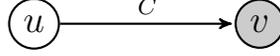
\begin{figure}[!h]
    \centering
        \begin{tikzpicture}[->,>=stealth',shorten >=1pt,auto,node distance=3cm,
  thick,main node/.style={circle,fill=white!20,draw,font=\sffamily\Large\bfseries}]
  \node[main node] (2) [central_node_fill] {$v$};
  \node[main node] (1) [left of=2]{$u$};
  \path[every node/.style={font=\sffamily\small}]
		(1) edge node [above] {$C$} (2);
		(2)
\end{tikzpicture}
    \caption{Granger-causality graph. The variable $u$ Granger-causes the variable $v$ for the set of companies $C$.}
		\label{fig:G}
\end{figure}

\subsection{Predictive Analysis}

To evaluate the predictive power of sentiment we consider two auto-regressive models, with and without sentiment, and conduct a 1-step ahead prediction analysis:
\begin{align}
\mathcal{M}_0&:X(t) = {\alpha} + \sum^k_{\tau=1}{{\beta}_{\tau} X(t-\tau)} + \epsilon_t, & \label{eq:AR1}\\
\mathcal{M}_1&:X(t) = {\alpha} + \sum^k_{\tau=1}{{\beta}_{\tau} X(t-\tau)}+  \sum^k_{i=1}{{\gamma}_iY(t-\tau)}+ \widehat{\epsilon}_t \label{eq:AR2}
\end{align}
where,
\begin{align}
X(t) &\in \{ER(t), VOL(t)\},\\
Y(t) &\in \{G(t), B(t), SR(t), V(t)\}.
\end{align}
As the absolute sentiment $SA(t)$ (of both News and Twitter) is already a linear combination between positive $G(t)$ and negative stories $B(t)$, we will not consider it in the linear regression for any dataset. Moreover, we consider only 1 day lag for the sentiment variables and a lag of 2 days for the financial variables\footnote{A model selection approach can be also used in order to find an optimal lag for the explanatory variables, examples of selection's criteria are: the Akaike information criterion (AIC), the Bayesian information criterion (BIC) and Mallow's Cp. See \cite{box1976time}.}. Again, we will consider the normally standardized versions of the time series analysed. 

Hence, we will consider the following regression models for the excess of log-return prediction:
\begin{align}
\mathcal{M}_0:ER(t) =\ & \ \alpha + \beta_1ER(t-1) + \beta_2ER(t-2) + \epsilon_t,  \label{eq:AR1news}\\
\mathcal{M}_1:ER(t) =\ & \ \alpha + \beta_1ER(t-1) + \beta_2ER(t-2)   \\
\ &+\gamma_1G(t-1)+\gamma_2B(t-1)+\gamma_3SR(t-1)+ \widehat{\epsilon}_t \nonumber
\end{align}

For the volatility prediction using news as datasource we will not include the volume time series $V_{News}(t)$ as an explanatory variable in the regression because of its high correlation with the number of positive and negative news already taken into account in the model. Notice that, for the Twitter case, the volume time series consider also non-English messages which are not taken into account by the time series in $G_{Twitter}(t)$ and $B_{Twitter}(t)$. Therefore, we keep $V_{Twitter}(t)$ as an explanatory variable in the Twitter model. As a result, we have for news:
\begin{align}
\mathcal{M}_0:VOL(t) =\ &{\alpha} + \beta_1VOL(t-1) + \beta_2VOL(t-2) + \epsilon_t, \\
\mathcal{M}_1:VOL(t) =\ &{\alpha} + \beta_1VOL(t-1) + \beta_2VOL(t-2)\\
\ & + \gamma_1G_{News}(t-1)+\gamma_2B_{News}(t-1) + \widehat{\epsilon}_t  \nonumber
\end{align}
and for Twitter:
\begin{align}
\mathcal{M}_0:VOL(t) =\ &{\alpha} + \beta_1VOL(t-1) + \beta_2VOL(t-2) + \epsilon_t,\\
\mathcal{M}_1:VOL(t) =\ &{\alpha} + \beta_1VOL(t-1) + \beta_2VOL(t-2) \\
\ & + \gamma_1G_{Twitter}(t-1)+\gamma_2B_{Twitter}(t-1)+\gamma_3V_{Twitter}(t-1)+ \widehat{\epsilon}_t. \nonumber 
\end{align}

Forecasting accuracy is measured by comparing the two residuals $\epsilon_t$ and $\widehat{\epsilon}_t$ in terms of Residual Standard Error:
\begin{align}
\hat{\sigma} = \sqrt{\frac{\sum\limits^{T}_{i=1}{\left(y_i-\widehat{y}_i\right)^2}}{n}} = \sqrt{\frac{\sum\limits^{T}_{i=1}{\hat{\epsilon}^2_i}}{n}}
\end{align}
where, $T$ is the total number of points, $n$ is the number of degrees of freedom of the model, $\widehat{y}_i$ are the predicted values and $y_i$ are the observed ones. 


\section{Results and Discussion}
Here we present the results from the Granger-causality tests and the predictive analysis between the financial variables and sentiment data from Twitter and news. The sentiment predictive power and its Granger-causality tests are fulfilled in 1-step ahead fashion. We investigate the statistical significance of the sentiment variables in regards to movements in returns and volatility and we compare the Twitter results with news. We provide empirical evidence that Twitter is moving the market in respect to the excess of log-returns for a subset of stocks. Also, Twitter presents a stronger relationship with stock returns compared to news, in the selected retail companies. On the other hand, the Twitter sentiment analytics showed a weaker relationship with volatility compared to news.

\subsection{Excess of Log-Returns}

We analyse the dynamics of excess of log-returns of the stocks considered in relation to absolute and relative sentiments and also with the number of positive and negative stories.

Fig. \ref{fig:G1} shows the Granger-causality graph that summarises the significant Granger-causalities (p-value $< 0.05$) between the excess of log-return and the sentiment variables for both news and Twitter. See Table \ref{tb:GRANGER-ER} for the detailed results. We observe that the Twitter's sentiment analytics present more significant points compared to news. The Twitter's relative sentiment and its number of positive messages Granger-cause log-returns, respectively, for the companies GAMESTOP CORP. and MATTEL INC. The Twitter's absolute sentiment also Granger-causes returns for MATTEL INC. while having a two-way significant (p-value $<0.01$) Granger-causality for the company HOME DEPOT INC. The number of negative stories alone has no significant relationship with returns but combined with the number of positive stories in the form of relative and absolute sentiment it shows to be an important measure. The News' analytics have only one significant relationship that it is observed in the number of positive news Granger-causing the excess of log-returns for the company GAMESTOP CORP.

\tikzstyle{central_node_fill}=[circle,fill=black!20,draw,font=\sffamily\Large\bfseries]

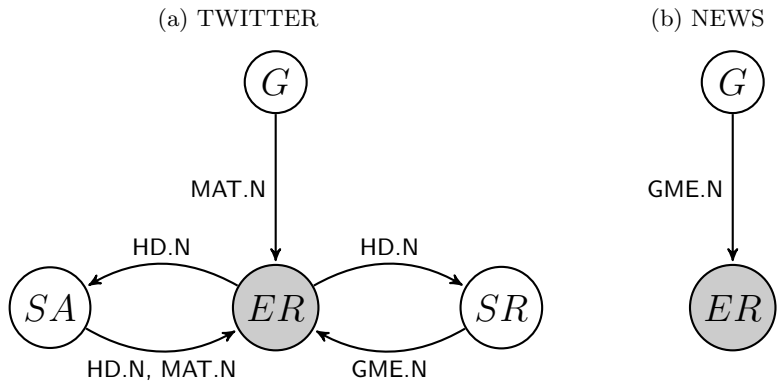
\begin{figure*}[!h]
    \centering
    \begin{subfigure}[t]{0.5\textwidth}
        \centering
				\caption{TWITTER}
        \begin{tikzpicture}[->,>=stealth',shorten >=1pt,auto,node distance=3cm,
  thick,main node/.style={circle,fill=white!20,draw,font=\sffamily\Large\bfseries}]
  \node[main node] (5) [central_node_fill] {$ER$};
  \node[main node] (1) [above of=5]{$G$};
  \node[main node] (3) [right of=5] {$SR$};
  \node[main node] (4) [left of=5] {$SA$};
  \path[every node/.style={font=\sffamily\small}]
		(1) edge node [left] {MAT.N} (5) 
    (3) edge [bend left] node [below] {GME.N} (5)
    (4) edge [bend right] node [below] {HD.N, MAT.N} (5)
    (5) edge [bend left] node[above] {HD.N} (3)
    	edge [bend right] node[above] {HD.N} (4);
\end{tikzpicture}
    \end{subfigure}%
    ~ 
    \begin{subfigure}[t]{0.5\textwidth}
        \centering
				\caption{NEWS}
        \begin{tikzpicture}[->,>=stealth',shorten >=1pt,auto,node distance=3cm,
  thick,main node/.style={circle,fill=white!20,draw,font=\sffamily\Large\bfseries}]
  \node[main node] (5) [central_node_fill] {$ER$};
  \node[main node] (1) [above of=5]{$G$};
  \path[every node/.style={font=\sffamily\small}]
		(1) edge node [left] {GME.N} (5)
    (5) ;
\end{tikzpicture}    
    \end{subfigure}
    \caption{Granger-causality graph for (a) Twitter and (b) news. It shows the significant points in the the Granger-causality test between excess of log-returns ($ER$) and the sentiment analytics: number of positive stories ($G$), number of negative stories ($B$), absolute sentiment ($SA$) and relative sentiment $SR$. Sentiment variables that presented no significant causality are not shown in the graph.}
		\label{fig:G1}
\end{figure*}

The solution of the multiple regression analysis in Table \ref{tb:LR-ER} agrees with the Granger-causality tests as it shows Twitter with a higher number of significant sentiment coefficients compared to news. The company MATTEL INC. particularly presents all sentiment coefficients with high significance (p-value $<$ 0.01) suggesting that the Twitter sentiment analytics is indeed relevant in the prediction of the next day excess of log-return. The companies HOME DEPOT INC. and GAMESTOP CORP. also presented significant sentiment coefficients. For the News analytics, the sentiment was significant only for the company GAMESTOP CORP. Further, analysis of the Residual Standard Error of the models with and without sentiment variables in Table \ref{tb:RSE-ER} shows that the use of the Twitter sentiment variables reduced the error of the model with only market data for the companies MATTEL INC., HOME DEPOT INC. and GAMESTOP CORP. while the news' sentiment improved the prediction only for the company GAMESTOP CORP.

In sum, the Twitter analytics surprisingly showed a stronger causality with stock's returns compared to news. It is interesting to notice that we did not perform any explicit filtering process in the Twitter analytics. However, we only considered the extremes of polarity (positive and negative categories), i.e., we did not consider the neutral tweets. This suggests that the sentiment classification itself is indirectly filtering the noise in the data in the sense that the non-neutral tweets are really informative. Moreover, the increased causality compared to news indicates a prominent role of Twitter in the retail industry. We believe that Twitter act as a feedback channel for the retail brands in a timely fashion fine-grained way compared to News.

\begin{table}[!h]
\centering
\caption{Residual standard error improvement in prediction of excess of log-return $ER(t)$ using $SR(t), G(t)$ and $B(t)$ compared to the model with only market data.} 
\label{tb:RSE-ER}
\begin{center}
 \begin{adjustbox}{max width=\textwidth}
 \begin{tabular}{l c c}
& \multicolumn{2}{c}{Error Reduction (\%)} 
\\ \cline{2-3}				
Company & NEWS	& TWITTER	\\ \hline
NIKE INC. & -2.41	& -0.58	\\
ABERCROMBIE \& FITCH CO. & -1.26 & -0.60 \\
HOME DEPOT INC. & -0.99	 & 1.23\\ 
MATTEL INC. & -0.48 & 2.82\\
GAMESTOP CORP. & 8.34 & 1.10
\\
\hline
\end{tabular}
\end{adjustbox}
\end{center}
 \end{table}  \FloatBarrier

\begin{table}[!h]
\centering
\rotatebox{90}{
\begin{minipage}{\textheight}
\centering
\caption{Statistical significance (p-values) of Granger-causality analysis between excess log-return and $SA(t), SR(t), G(t)$ and $B(t)$ for the companies: ABERCROMBIE \& FITCH CO. (ANF.N), NIKE INC. (NKE.N), HOME DEPOT INC. (HD.N), MATTEL INC. (MAT.N) and GAMESTOP CORP. (GME.N).} 
\label{tb:GRANGER-ER}
 \begin{adjustbox}{max width=\textwidth}
 \begin{tabular}{l c c c c c c c c c c c}
& \multicolumn{5}{c}{TWITTER ANALYTICS} &  \multicolumn{5}{c}{NEWS ANALYTICS}\\ 
\cline{2-6} \cline{8-12}
& HD.N	& MAT.N	& GME.N	& NKE.N	& ANF.N & & HD.N	& MAT.N	& GME.N	& NKE.N	& ANF.N\\ \hline
$SA \rightarrow ER$	& 0.003***	& 0.046**	& 0.140	& 0.888	& 0.477 & &0.303	& 0.411	& 0.140	& 0.621	& 0.707\\ 
$ER \rightarrow SA$	& 0.006***	& 0.404	& 0.231	& 0.354	& 0.937 & &0.423	& 0.451	& 0.230	& 0.546	& 0.281\\
$SR \rightarrow ER$	& 0.449	& 0.497	& 0.032**	& 0.680	& 0.591 & &0.747	& 0.977	& 0.696	& 0.816	& 0.814\\ 
$ER \rightarrow SR$	& 0.024**	& 0.855	& 0.196	& 0.995	& 0.875 & &0.942	& 0.314	& 0.162	& 0.564	& 0.213\\ 
$G \rightarrow ER$	& 0.182	& 0.016**	& 0.885	& 0.400	& 0.685 & &0.203	& 0.228	& 0.014**	& 0.304	& 0.231\\ 
$ER \rightarrow G$	& 0.050*	& 0.305	& 0.957	& 0.380	& 0.197 & &0.388	& 0.382	& 0.171	& 0.199	& 0.518\\ 
$B \rightarrow ER$ & 0.327	& 0.559	& 0.267	& 0.344	& 0.763 & &0.681	& 0.976	& 0.920	& 0.796	& 0.398\\ 
$ER \rightarrow B$	& 0.219	& 0.792	& 0.538	& 0.166	& 0.480 & &0.855	& 0.646	& 0.894	& 0.769	& 0.863\\ 
\hline
\multicolumn{11}{c}{Significance codes: p-value $< 0.01$: ***, p-value $< 0.05$: **, p-value $< 0.1$: *}
\end{tabular}
\end{adjustbox}
\end{minipage}
}
 \end{table}  \FloatBarrier

\begin{table}[!h]
\centering
\rotatebox{90}{
\begin{minipage}{\textheight}
\caption{Summary statistics of multiple regression. Prediction of excess of log-return ($ER$) using sentiment ($SR, G, B$) for the companies: ABERCROMBIE \& FITCH CO. (ANF.N), NIKE INC. (NKE.N), HOME DEPOT INC. (HD.N), MATTEL INC. (MAT.N) and GAMESTOP CORP. (GME.N).} 
\label{tb:LR-ER}
 \begin{adjustbox}{max width=\textwidth}
 \begin{tabular}{l c c c c c c c c c c c}
& \multicolumn{5}{c}{TWITTER ANALYTICS} & \multicolumn{5}{c}{NEWS ANALYTICS}  \\ 
\cline{2-6} \cline{8-12}
 & HD.N & MAT.N & GME.N & NKE.N & ANF.N & & HD.N & MAT.N & GME.N & NKE.N & ANF.N\\ 
\cline{2-6} \cline{8-12}
\multicolumn{1}{l}{$ER_{(t-1)}$}  & 4.414018e-04  & 1.192513e-01* & -1.704137e-01** & -3.001408e-02  &  1.203011e-02 & &   2.592302e-01  &  6.129545e-01*** &  -9.060641e-02 & 4.018227e-02  &   -1.457410e-01 \\ 
\multicolumn{1}{l}{$ER_{(t-2)}$}  & -9.806531e-02  &  9.192829e-03 &  1.535305e-02 &   9.752118e-02  &  -4.741730e-02 &   &   -5.695452e-02  &  -2.674834e-01*&  4.458446e-01*** &4.363861e-02  &   -6.310047e-02 \\ 
\multicolumn{1}{l}{$SR_{(t-1)}$}   &  -2.258381e-02 & 3.718939e-01*** & 1.821209e-01**&   1.817089e-02  & 2.558886e-02&   &   -1.092573e-01  &   -8.211806e-02  &   -2.852138e-01  & 4.487921e-02  &   9.364799e-02\\ 
\multicolumn{1}{l}{$G_{(t-1)}$}   & 2.536661e-01*** & -4.286924e-01*** &  -1.229142e-02 &   -3.956882e-02  & 4.445273e-02&   &   2.600786e-01  &   2.683852e-01  &  4.962514e-01***&-2.144345e-01  &   -2.279505e-01\\
\multicolumn{1}{l}{$B_{(t-1)}$}   & -2.612111e-01** & 4.801024e-01*** & -1.159965e-01&   -1.459096e-02  & -3.347246e-02&   &   -9.858689e-02  &  3.625886e-03  &   -5.315150e-02&-6.179474e-02  &   3.457372e-02   \\ \hline
\multicolumn{11}{c}{Significance codes: p-value $< 0.01$: ***, p-value $< 0.05$: **, p-value $< 0.1$: *} \\ 
\end{tabular}
\end{adjustbox}
\end{minipage}
}
 \end{table}  \FloatBarrier

\subsection{Volatility}

Here, we analyse the interplay between message's volume and sentiment with stock's volatility. As volume measures we consider: the number of positive and negative English stories and also the total volume of stories, regardless of the language. As sentiment analytics we consider: absolute sentiment and relative sentiment.

Fig. \ref{fig:G2} shows the significant links (p-value $< 0.05$) of the Granger-causality test between the volatility and the sentiment variables. See Table \ref{tb:GRANGER-VOL} for the detailed results. Overall, there are more significant points of causality for the News sentiment analytics compared to Twitter. We observe that the number of positive stories and the total volume both Granger-cause volatility for Twitter as well as for news but more companies are affected by news. The absolute sentiment Granger-causes volatility only for news, observed in the company ABERCROMBIE \& FITCH CO. (ANF.N). The relative sentiment and the number of negative stories are not causing volatility, on the other hand volatility is Granger-causing negative news for the company GAMESTOP CORP. (GME.N).

\tikzstyle{central_node_fill}=[circle,fill=black!20,draw,font=\sffamily\Large\bfseries]

\begin{figure*}[!h]
    \centering
    \begin{subfigure}[t]{0.5\textwidth}
        \centering
				\caption{TWITTER}
        \begin{tikzpicture}[->,>=stealth',shorten >=1pt,auto,node distance=3cm,
  thick,main node/.style={circle,fill=white!20,draw,font=\sffamily\Large\bfseries}]
  \node[main node] (5) [central_node_fill] {$VOL$};
  \node[main node] (1) [below of=5]{$G$};
  \node[main node] (2) [above of=5] {$V$};
  
  \path[every node/.style={font=\sffamily\small}]
(1) edge node [left] {HD.N} (5) 
(2) edge [bend left] node [right] {HD.N, MAT.N} (5) 
    (5) edge [bend left] node[left] {MAT.N} (2);
\end{tikzpicture}
    \end{subfigure}%
    ~ 
    \begin{subfigure}[t]{0.5\textwidth}
        \centering
				\caption{NEWS}
        \begin{tikzpicture}[->,>=stealth',shorten >=1pt,auto,node distance=3cm,
  thick,main node/.style={circle,fill=white!20,draw,font=\sffamily\Large\bfseries}]
  \node[main node] (5) [central_node_fill] {$VOL$};
  \node[main node] (1) [below of=5]{$G$};
  \node[main node] (2) [above of=5] {$V$};
  \node[main node] (3) [left of=5]{$SA$};
  \node[main node] (4) [right of=5] {$B$};
  
  \path[every node/.style={font=\sffamily\small}]
(1) edge node [left] {GME.N, ANF.N} (5) 
(2) edge node [left] {HD.N, GME.N, ANF.N} (5)
    (3) edge node [above] {ANF.N} (5)
    (4)
    (5) edge node[above] {GME.N} (4);
\end{tikzpicture}    
    \end{subfigure}
    \caption{Granger-causality graph for (a) Twitter and (b) news. It shows the significant points in the Granger-causality test between volatility ($VOL$) and the sentiment analytics: total number of stories ($V$), number of positive stories ($G$), number of negative stories ($B$), absolute sentiment ($SA$) and relative sentiment ($SR$). Sentiment variables with no significant causality are not shown in the graph. }
		\label{fig:G2}
\end{figure*}
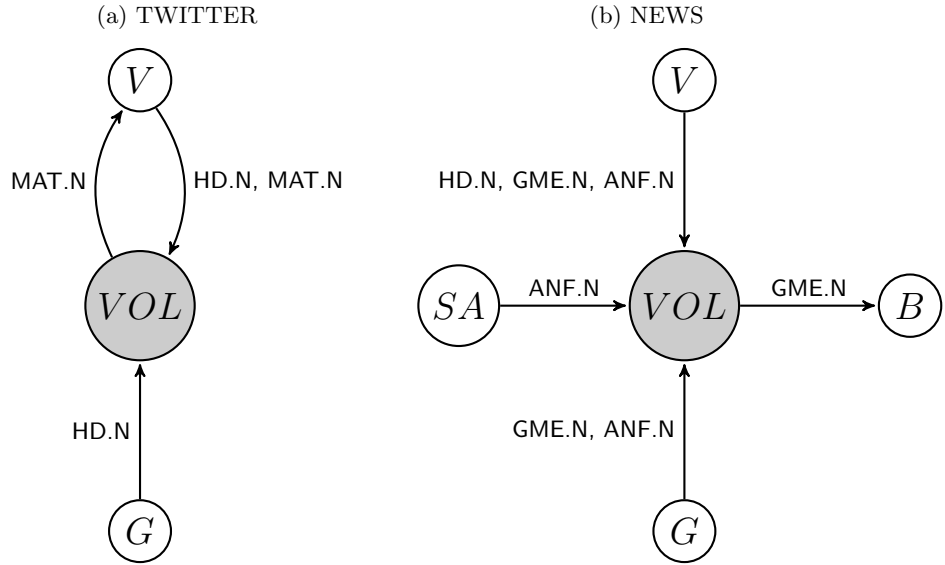

The solution of the multiple regression analysis in Table \ref{tb:LR-VOL} shows that the number of positive stories 
is a significant variable for both news and Twitter, being more significant for the former. The number of negative stories showed no relevance in both regressions. The total volume of Twitter was relevant only for the company NKE.N. Moreover, analysis of the Residual Standard Error of the models with and without sentiment variables in Table \ref{tb:RSE-VOL} shows that both Twitter and News were able to reduce the error in prediction for a subset of the companies. In the cases where the model was improved with sentiment, News provided a higher reduction of error compared to Twitter. 

Overall, the news analytics showed a higher causality with volatility compared to Twitter. This confirms the predictive power of News with volatility as described in the literature \cite{diBartolomeo:2009, Mitra:2013, Sidorov:2014, Kalev:2011}. Further improvements in entity detection in the sentiment classification algorithm used in the dataset provided \cite{1507.00955} may improve the Twitter's results.

\begin{table}[H]
\centering
\caption{Residual standard error improvement in the prediction of volatility $VOL(t)$ using $G(t), B(t)$ and $V(t)$ compared to the model with only market data.} 
\label{tb:RSE-VOL}
\centering
 \begin{adjustbox}{max width=\textwidth}
 \begin{tabular}{l c c}
& \multicolumn{2}{c}{Error Reduction (\%)} 
\\ \cline{2-3}				
Company & NEWS	& TWITTER	\\ \hline
NIKE INC. & 1.36	& 1.08	\\
ABERCROMBIE \& FITCH CO. & 4.03 & -0.52 \\
HOME DEPOT INC. & 2.46	 & 1.10\\ 
MATTEL INC. & -2.21 & -0.36\\
GAMESTOP CORP. & 14.99 & 0.20
\\
\hline
\end{tabular}
\end{adjustbox}
 \end{table}  \FloatBarrier

   \begin{table}[!h]
\centering
\rotatebox{90}{
\begin{minipage}{\textheight}
\caption{Statistical significance (p-values) of Granger-causality analysis between volatility ($VOL(t))$ and $G(t), B(t), SA(t), SR(t)$ and $V(t)$ for the companies: ABERCROMBIE \& FITCH CO. (ANF.N), NIKE INC. (NKE.N), HOME DEPOT INC. (HD.N), MATTEL INC. (MAT.N) and GAMESTOP CORP. (GME.N).} 
\label{tb:GRANGER-VOL}
 \begin{adjustbox}{max width=\textwidth}
 \begin{tabular}{l c c c c c c c c c c c}
& \multicolumn{5}{c}{TWITTER ANALYTICS} &  \multicolumn{5}{c}{NEWS ANALYTICS}\\ 
\cline{2-6} \cline{8-12}
& HD.N	& MAT.N	& GME.N	& NKE.N	& ANF.N & & HD.N	& MAT.N	& GME.N	& NKE.N	& ANF.N\\ \hline
$V \rightarrow VOL$	& 0.025**	& 0.004***	& 0.751& 0.976	& 0.729	&& 0.029**	& 0.320	& 0.020**	& 0.846	& 0.001***\\
$VOL \rightarrow V$	& 0.228	& 0.016**	& 0.980 & 0.611	& 0.924	&& 0.307	& 0.453	& 0.053*	& 0.462	& 0.560 \\
$G \rightarrow VOL$	& 0.004***	& 0.072*	& 0.678& 0.397	& 0.228	&& 0.060*	& 0.341	& 0.031**	& 0.961	& $<$0.001***  \\
$VOL \rightarrow G$		& 0.560	& 0.273	& 0.668& 0.330	& 0.690	&& 0.146	& 0.390	& 0.211	& 0.859	& 0.615\\
$B \rightarrow VOL$ & 0.053*	& 0.118	& 0.720& 0.636	& 0.884	&& 0.522	& 0.301	& 0.204	& 0.526	& 0.540\\
$VOL \rightarrow B$		& 0.420 & 0.477	& 0.363	& 0.729	& 0.967	&& 0.484	& 0.620	& 0.026**	& 0.391	& 0.801\\
$SR \rightarrow VOL$	& 0.539	& 0.305	& 0.489& 0.786	& 0.220	&& 0.416	& 0.925	& 0.510	& 0.441	& 0.056*\\
$VOL \rightarrow SR$	& 0.944	& 0.867	& 0.791& 0.611	& 0.623	&& 0.854	& 0.352	& 0.590	& 0.184	& 0.340\\
$SA \rightarrow VOL$	& 0.274	& 0.736	& 0.437& 0.142	& 0.216	&& 0.435	& 0.194	& 0.497	& 0.652	& 0.010**\\
$VOL \rightarrow SA$ & 0.458	& 0.900	& 0.747 & 0.120	& 0.950 && 0.173	& 0.352	& 0.187	& 0.707	& 0.996\\
\hline
\multicolumn{11}{c}{Significance codes: p-value $< 0.01$: ***, p-value $< 0.05$: **, p-value $< 0.1$: *}
\end{tabular}
\end{adjustbox}
\end{minipage}
}
 \end{table}  \FloatBarrier

  \begin{table}[!h]
\centering
\rotatebox{90}{
\begin{minipage}{\textheight}
\caption{Summary statistics of multiple regression. Prediction of volatility ($VOL$) using sentiment ($G, B$) and volume ($V$) for the companies: ABERCROMBIE \& FITCH CO. (ANF.N), NIKE INC. (NKE.N), HOME DEPOT INC. (HD.N), MATTEL INC. (MAT.N) and GAMESTOP CORP. (GME.N).} 
\label{tb:LR-VOL}
 \begin{adjustbox}{max width=\textwidth}
 \begin{tabular}{l c c c c c c c c c c c}
&  \multicolumn{5}{c}{TWITTER ANALYTICS} & \multicolumn{5}{c}{NEWS ANALYTICS} \\ 
\cline{2-6} \cline{8-12}
 & HD.N & MAT.N & GME.N & NKE.N & ANF.N&  & HD.N & MAT.N & GME.N & NKE.N & ANF.N \\ 
\cline{2-6} \cline{8-12}
\multicolumn{1}{l}{$VOL_{(t-1)}$}  & 2.305851e-01*** & 2.211374e-01*** &  1.019004e-01&  1.893666e-01***& 1.308395e-01*& &3.891e-01* & 6.265e-02& 9.587e-02 & 3.785e-01**& 3.792e-01**\\ 
\multicolumn{1}{l}{$VOL_{(t-2)}$}   & 1.250733e-01* & 1.238580e-01* &  -6.173133e-02&  4.307688e-02  & 8.735957e-02&&-9.120e-02 & -4.451e-03& 5.092e-02&3.093e-02 & 1.054e-01\\ 
\multicolumn{1}{l}{$G_{(t-1)}$}   & 1.845026e-01* &  1.073941e-01 &  -2.807417e-01*& -1.586209e-03 & 3.533443e-02&&3.018e-01** & 2.772e-03& 6.046e-01***& 1.274e-01  & -3.329e-01*\\ 
\multicolumn{1}{l}{$B_{(t-1)}$}   &  1.420223e-03 &  -7.759929e-02 &  4.461084e-02&  -2.181354e-03  & -8.569212e-02&& 9.294e-02& -1.105e-01& 1.184e-01&-1.400e-01 & -1.300e-01 \\ 
\multicolumn{1}{l}{$V_{(t-1)}$}  & -1.468387e-03  & -1.862391e-02 & 1.920558e-01& 1.901394e-01** & 5.742508e-02&  & N/A & N/A & N/A & N/A  & N/A\\ \hline
\multicolumn{11}{c}{Significance codes: p-value $< 0.01$: ***, p-value $< 0.05$: **, p-value $< 0.1$: *}
\end{tabular}
\end{adjustbox}
\end{minipage}
}
 \end{table}  \FloatBarrier

\section{Conclusion}

We  showed  that  measures of the Twitter sentiment extracted from listed retail brands have statistically-significant relationship with stock returns and volatility. While analysing the interplay between the excess of log-return and the Twitter sentiment variables we conclude that: (i) Twitter presented a stronger Granger-causality with stock's returns compared to news; (ii) positive tweets and Twitter's sentiment Granger-cause excess of log-returns for a subset of companies; (iii) Twitter's sentiment analytics is indeed relevant in the prediction of the next day excess of log-return even compared to traditional newswires. Moreover, in the volatility analysis we found that: (i) Twitter's analytics showed a weaker relationship with volatility compared to the one observed with returns; (ii) the number of positive tweets and the total volume both Granger-cause volatility for some companies but present reduced Granger-causality compared to news; (iii) the number of positive tweets is a significant variable for the 1-step ahead prediction of volatility while the number of negative messages showed no relevance. Overall, the Twitter sentiment analytics showed to be a distinct and complementary proxy of market's sentiment compared to news in the analysis of the financial dynamics of retail brands' stocks. Surprisingly, the Twitter's sentiment presented a relatively stronger relationship with the stock's returns compared to traditional newswires. The results suggest that social media analytics have a prominent role in the dynamics of the retails sector in the financial markets.

\section*{Acknowledgments}

We thank the valuable feedback from the two anonymous reviewers. This work was supported by OptiRisk Systems which provided the news sentiment analytics from RavenPack. T.A. acknowledges support of the UK Economic and Social Research Council (ESRC) in funding the Systemic Risk Centre (ES/K002309/1). T.T.P.S. acknowledges financial support from CNPq - The Brazilian National Council for Scientific and Technological Development. O.K. acknowledges support from the company Certona Corporation.
\bibliographystyle{splncs03}
\bibliography{chapter}

\end{document}